\documentclass{ieeeaccess}
\usepackage{cite}
\usepackage{amsmath,amssymb,amsfonts}
\usepackage{algorithm}
\usepackage{algorithmicx}
\usepackage{algpseudocode}
\usepackage{graphicx}
\usepackage{textcomp}
\usepackage{multirow}
\usepackage{color}
\floatname{algorithm}{Algorithm}

\begin{document}
\history{Date of publication xxxx 00, 0000, date of current version xxxx 00, 0000.}
\doi{10.1109/ACCESS.2020.DOI}

\title{Spatiotemporal Differences of COVID-19 Infection among Healthcare Workers and Patients in China from January to March 2020}
\author{\uppercase{Peixiao Wang}\authorrefmark{1},
\uppercase{Xinyan Zhu\authorrefmark{1,3,4},
Wei Guo\authorrefmark{1},
Hui Ren\authorrefmark{1},
Tao Hu}\authorrefmark{2}}
\address[1]{State Key Laboratory of Information Engineering in Surveying, Mapping and Remote Sensing, Wuhan University, Wuhan 430079, China;  (e-mail: peixiaowang@whu.edu.cn,guowei-lmars@whu.edu.cn,renhui@whu.edu.cn) }
\address[2]{Center for Geographic Analysis, Harvard University, Cambridge, MA 02138, USA}
\address[3]{Collaborative Innovation Center of Geospatial Technology, Wuhan 430079, China}
\address[4]{Key Laboratory of Aerospace Information Security and Trusted Computing, Ministry of Education, Wuhan University, Wuhan 430079, China}
\tfootnote{This project was supported by the Key Program of National Natural Science Foundation of China (No. 41830645), National Key Research and Development Program of China (Grant Nos. 2018YFB0505500,2018YFB0505503), Funding program: CAE Advisory Project No. 2020-ZD-16.}

\markboth
{PEIXIAO WANG \headeretal: Preparation of Papers for IEEE ACCESS}
{PEIXIAO WANG \headeretal: Preparation of Papers for IEEE ACCESS}

\corresp{Corresponding author: Xinyan Zhu (e-mail: xinyanzhu@whu.edu.cn) and Tao Hu (e-mail: taohu@g.harvard.edu)}

\begin{abstract}
Studying the spatiotemporal differences in coronavirus disease (COVID-19) between social groups such as healthcare workers (HCWs) and patients can aid in formulating epidemic containment policies. Most previous studies of the spatiotemporal characteristics of COVID-19 were conducted in a single group and did not explore the differences between groups. To fill this research gap, this study assessed the spatiotemporal characteristics and differences among patients and HCWs infection in Wuhan, Hubei (excluding Wuhan), and China (excluding Hubei). The temporal difference was greater in Wuhan than in the rest of Hubei, and was greater in Hubei (excluding Wuhan) than in the rest of China. The incidence was high in healthcare workers in the early stages of the epidemic. Therefore, it is important to strengthen the protective measures for healthcare workers in the early stage of the epidemic. The spatial difference was less in Wuhan than in the rest of Hubei, and less in Hubei (excluding Wuhan) than in the rest of China. The spatial distribution of healthcare worker infections can be used to infer the spatial distribution of the epidemic in the early stage and to formulate control measures accordingly.
\end{abstract}

\begin{keywords}
Healthcare worker infection; patient infection; spatiotemporal distribution; spatiotemporal differences; COVID-19
\end{keywords}

\titlepgskip=-15pt

\maketitle

\section{Introduction}
\PARstart{T}{he} COVID-19 virus was first discovered at the end of 2019, and subsequently spread rapidly worldwide. At present, the world has been heavily hit by the COVID-19 pandemic \cite{b1,b2}. Since the outbreak of COVID-19, many researchers have performed classical, epidemiological, mathematical, and statistical analyses to carry out emergency research from the perspectives of pathology \cite{b3,b4}, epidemiology \cite{b5}, genomics \cite{b6}, clinical medicine \cite{b7,b8,b9}, and molecular biology \cite{b10}. However, when a novel infectious disease is prevalent, relevant preventative measures such as self-isolation, restriction of crowd movement and gathering, and wearing of masks became important emergency containment measures \cite{b11,b12,b13}. 

To achieve targeted and accurate containment, it is necessary to understand the spatiotemporal spread of COVID-19 as time progresses and determine its development trend early-on \cite{b14}. In the early stages of the epidemic, patient infection data were a very important data source. Patient infection data contains a wealth of information that can be explored via spatiotemporal analysis \cite{b15,b16}. Therefore, many researchers have used this data to explore the spatiotemporal characteristics of COVID-19. However, studies using other data sources to assess the characteristics of the epidemic in social groups (e.g., healthcare workers) are scarce \cite{b17}. Studying the spatiotemporal distributions and differences in healthcare worker infections can aid in better formulating epidemic containment policies; however, the spatiotemporal characteristics of COVID-19 between healthcare workers are rarely studied. To fill this research gap, the spatiotemporal characteristics and differences between healthcare worker infections and patient infections were analyzed from the perspectives of Wuhan, Hubei (excluding Wuhan), and China (excluding Hubei) by combining the data of healthcare workers infection and the data of patients infection. The main contributions of this article are as follows:

(1) The spatiotemporal distributions of healthcare worker infections were analyzed from the perspectives of Wuhan, Hubei (excluding Wuhan), and China (excluding Hubei).

(2) The spatiotemporal differences among healthcare worker infections and patient infections were also analyzed. In addition, the study can provide a scientific reference for relevant departments to formulate epidemic containment policies from a new perspective.

\section{Related Works}

Many scientists have researched COVID-19 from different perspectives. In this section, we primarily present a review of relevant studies that explored the spatiotemporal distributions of COVID-19.

To explore the spatiotemporal distribution of the disease is mainly to summarize the spread law of COVID-19 to provide the basis for prevention and control. In the early stage of the epidemic, many studies were conducted based on patient confirmed data. For example, Zhang et al. \cite{b18} analyzed the epidemiological characteristics of COVID-19 in February 2020 based on patient confirmed data and reported the spatial distribution of COVID-19 early-on in China. Hu et al. \cite{b19} collected global patient confirmed data from different scales and established a data repository to provide data support for research related to the spatiotemporal distribution of COVID-19. Wang et al. \cite{b20} used spatiotemporal scanning statistics to detect the hotspots of new cases every week, based on the confirmed cases of COVID-19 at the county level in the United States, thereby characterizing the incidence trend. Further, Zhang et al. \cite{b21} compared the spatiotemporal characteristics between COVID-19 and SARS at the provincial level, thereby revealing the spread of COVID-19. Sun et al. \cite{b22} analyzed the regional spatiotemporal changes of the epidemic based on the daily new cases in 250 regions of South Korea, thereby evaluating South Korea’s containment strategy. 

Researchers have also combined patient confirmed data with other data sources to explore the spatiotemporal spread of COVID-19. For example, Zhang et al.\cite{b23} examined the factors influencing the number of imported cases from Wuhan and the spread speed and pattern of the pandemic by combining national flight and high-speed rail data. Based on mobile phone and confirmed patient data, Jia et al. \cite{b24} developed a spatiotemporal “risk source” model to determine the geographic distribution and growth pattern of COVID-19 and quickly as well as accurately assess the related risk. Zhang et al. \cite{b25} used the GeoDetector and the decision tree model to identify the main factors in low- and high-risk areas by combining multi-source data. Loske \cite{b26} explored the relationship between COVID-19 spread and transportation volume in food retail logistics by combining transport volume data and confirmed patient data. Further, Zhang et al. \cite{b27} measured the imported case risk of COVID-19 from inbound international flights by combining both daily dynamic international air connectivity data and updated global COVID-19 data.

However, the abovementioned studies are mostly based on patient confirmed data. The spatiotemporal distributions of COVID-19 in other groups are rarely explored. Thus, we analyzed the spatiotemporal characteristics of and differences between healthcare worker infections and patient infections from the perspectives of Wuhan, Hubei (excluding Wuhan), and China (excluding Hubei).

\section{Study Area and Data Sources}

\subsection{Study Area}
As of July 2020, the total number of locally confirmed COVID-19 cases in China had exceeded 80,000. Of these cases, 68,135 were confirmed in Hubei Province, among which 50,340 were confirmed in Wuhan, accounting for 81.5\% and 60.80\% of the country’s cases, respectively. To study the temporal and spatial spread of COVID-19 in China, the study area was divided into three parts: Wuhan City, Hubei Province (excluding Wuhan), and the rest of China. As shown in Figure 1, Hubei Province is in the central region of China, and Wuhan City is in the central region of Hubei Province.

\Figure[t!](topskip=0pt, botskip=0pt, midskip=0pt){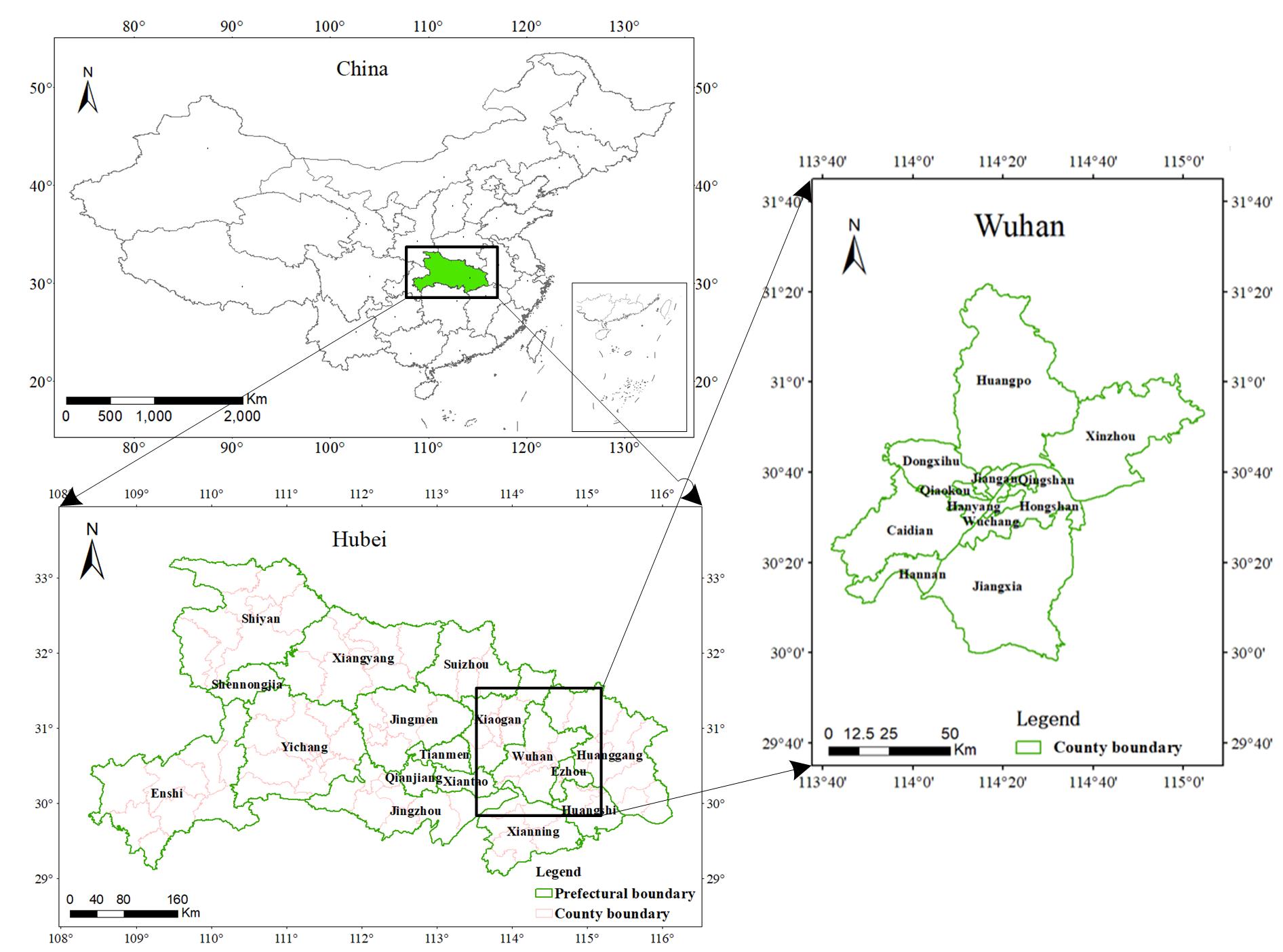}
{Sketch map of the study area.\label{fig1}}

\subsection{Data Sources and Data Preprocessing}
\subsubsection{Data Sources}

The data used in this study were divided into two categories: \emph{Patient Infection Table} and healthcare worker’s infection information. The \emph{Patient Infection Table} was mainly obtained from the National Health Commission of the People's Republic of China (http://www.nhc.gov.cn/xcs/yqtb/list\_gzbd.shtml) and spanned from January 15, 2020, to June 28, 2020. The confirmed information on healthcare workers was obtained mainly from the Chinese Red Cross Foundation (https://www.crcf.org.cn/article/20672). The Chinese Red Cross Foundation distributes relief funds to every healthcare worker with a confirmed case. As of July 31, 2020, 81 batches of confirmed healthcare workers had received foundation assistance. We first used Python language and web crawler technology to obtain 3,741 reports on confirmed healthcare workers from the Chinese Red Cross Foundation and later used the Baidu API for address matching. After matching their address, details of the city, province, and county of each healthcare worker with a confirmed case were obtained. The data format is shown in Table 1. Each record shows the confirmed time, province, city, and county of each confirmed healthcare worker.

\begin{table}
\begin{center}
\caption{Sample of healthcare worker infection data.}
\begin{tabular}{ccccc}
\hline
User ID&Date&Province&City&County\\
\hline
1&2020-01-15&Hubei&Wuhan&Huangpo\\
2&2020-01-15&Hubei&Jingmen&Zhongxiang\\
3&2020-02-04&Shandong&Qingdao&Shinan\\
...&...&...&...&...\\
3741&2020-02-01&Beijing&Beijing&Xicheng\\
\hline
\end{tabular}
\end{center}
\end{table}
\subsubsection{Data Preprocessing}

The \emph{Patient Infection Table} is collected and processed by a team at the China Data Center and is shared on the dataverse platform of Harvard University \cite{b19,b28}. Therefore, the \emph{Patient Infection Table} does not require additional data preprocessing. In this study, we mainly conducted data preprocessing for the confirmed case information of healthcare workers.

\begin{figure}[h]
	\centerline{\includegraphics[width=7.0cm]{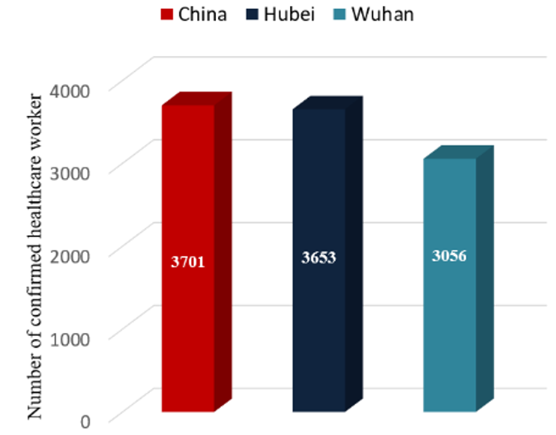}}
	\caption{Characteristics of healthcare worker infection data.}
	\label{fig}
\end{figure}

\Figure[t!](topskip=0pt, botskip=0pt, midskip=0pt)[width=17.0cm]{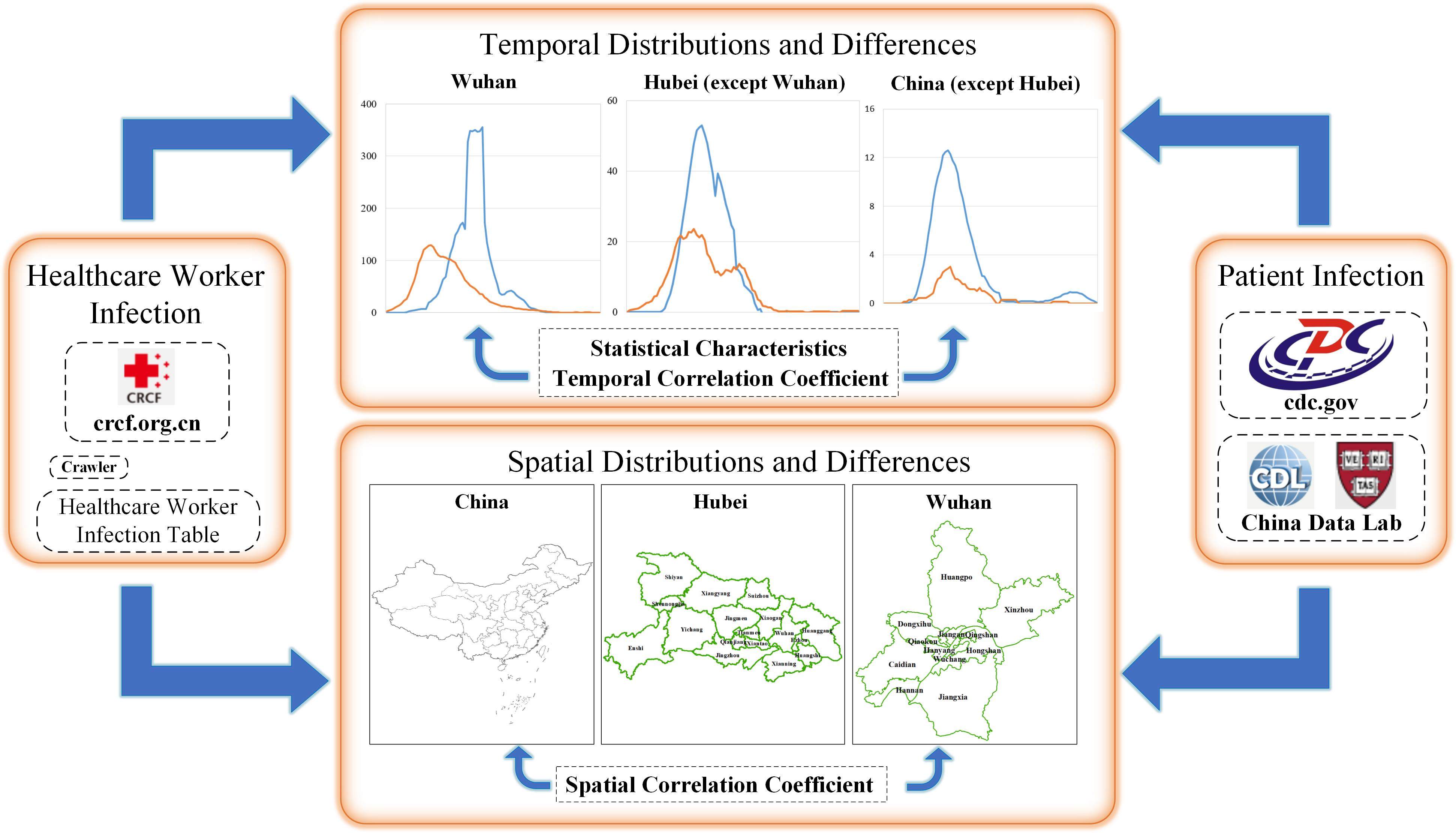}
{Technical framework of spatiotemporal distribution and differences in healthcare workers and patient infection.\label{fig3}}

Infection information on healthcare workers is mainly reported to the China Red Cross Foundation in two ways \cite{b29}: (1) direct reporting by individuals to the Red Cross Foundation of China, and (2) collection by the person’s respective hospital which then reports the information to the Red Cross Foundation of China. After examination and approval by the Red Cross Foundation of China, the rescue information is published on the website, after which the hospital may review it. Those who fail to pass the review do not qualify to be rescued. Therefore, data records of unqualified persons should be deleted from the original data. Second, the Chinese Red Cross Foundation not only rescued the infected healthcare worker, but also the infected or diseased staff during the epidemic; such data records need to be deleted. As shown in Figure 2, after data preprocessing, the data of a total of 3,701 healthcare workers infected with COVID-19 remained, including 3,667 from Hubei Province and 3,060 from Wuhan City.

\section{Methods}

The overall framework of the study is shown in Figure 3. First, the \emph{Healthcare Worker Infection Tables} in China, Hubei, and Wuhan, which recorded a high number of healthcare workers infection each day, were constructed. Second, based on the \emph{Infection Tables}, relevant statistical indicators such as the mean, variance, change speed (rise and decline periods), and the inflection point were used to analyze the time spread characteristics of healthcare worker and patient infections; we used Pearson’s correlation coefficient to calculate the time correlation between healthcare worker infection and patient infection. Finally, the thematic map method was used to analyze the spatial spread characteristics of infections among healthcare workers and patients. We also calculated the spatial correlation between healthcare worker and patient infections.

\subsection{Construction of the Healthcare Worker Infection Table}

The healthcare worker infection data which were collected using the crawler record the basic information of confirmed healthcare workers, however, the data cannot visually display the temporal and spatial distributions and spread of the confirmed healthcare worker’s infections. Therefore, based on the modeling idea of "province - city - county," this study constructed \emph{Healthcare Worker Infection Tables} based on the Partition Statistics from three perspectives: China, Hubei, and Wuhan.

The data structure for the \emph{Healthcare Worker Infection Tables} are shown in Figure 4. The \emph{Table(S, T)} represents a spatiotemporal dataset, where S and T are the spatial and temporal dimensions, respectively; $S = \{s_1,s_2,· · · ,s_m \}$ represents the row index of the table, and $T = \{t_1,t_2,· · · ,t_n \}$ represents the column index of the table.

\Figure[t!](topskip=0pt, botskip=0pt, midskip=0pt)[width=6.0cm]{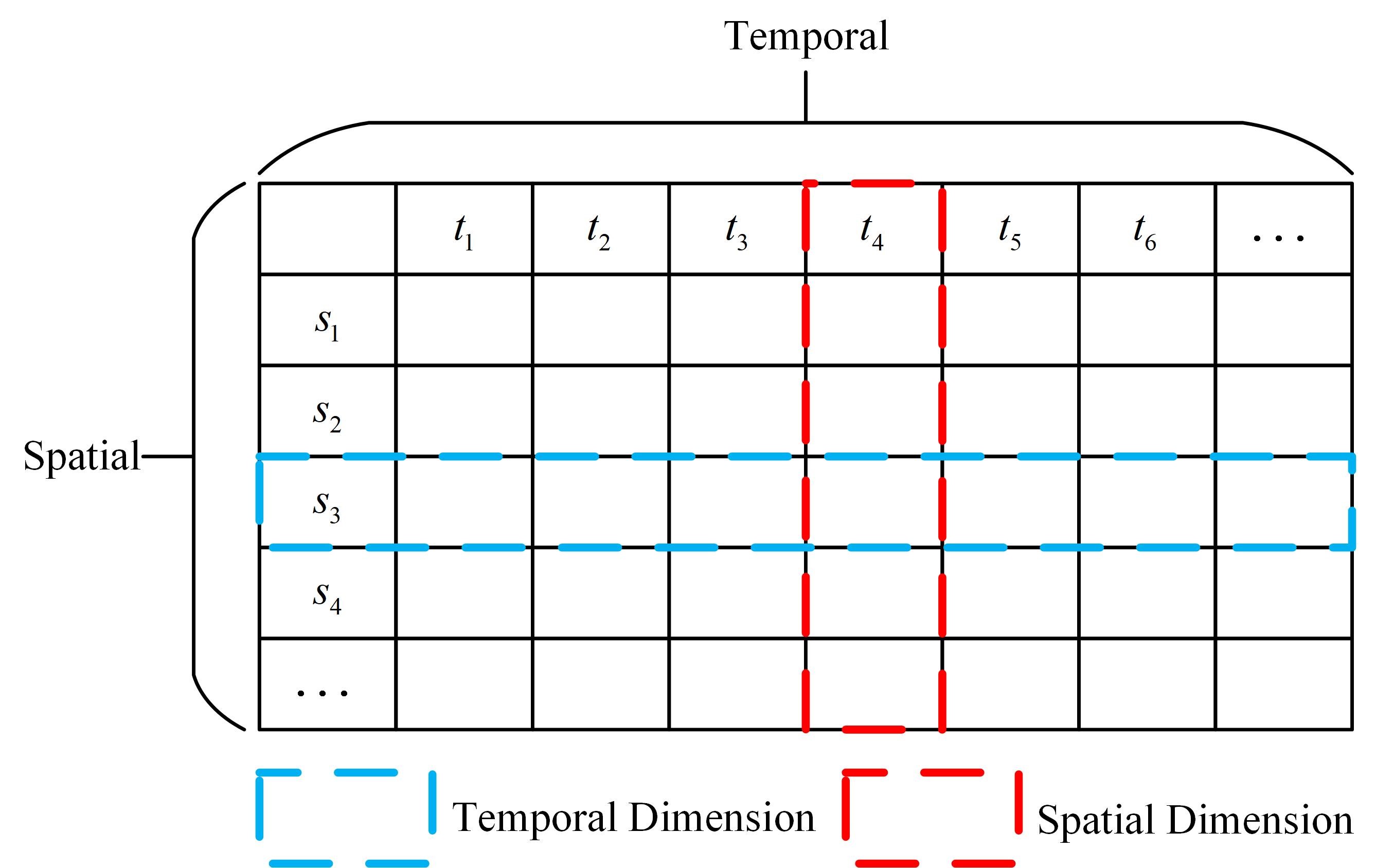}
{ Data structure of the confirmed inventories.\label{fig4}}

Taking the \emph{Healthcare Worker Infection Table} of Wuhan as an example, the construction process is shown in Algorithm 1. $\{cms_i\}$ represents healthcare worker infection data, and $Table^{Wuhan}$ represents the \emph{Healthcare Worker Infection Table} from the perspective of Wuhan. First, $Table^{Wuhan}$ is initialized to 0 (line 1). Then, the values in the table are assigned iteratively (lines 2-4), and $Table^{Wuhan}$ describes in detail the changes in healthcare worker infections in every county of Wuhan over time. Final, $Table^{Wuhan}$ is returned (line 5). The corresponding \emph{Healthcare Worker Infection Tables} in China and Hubei are expressed as $Table^{China}$ and $Table^{Hubei}$, respectively. The Healthcare Worker Infection Table of China records the number of healthcare worker infections every day in each province of China, and the \emph{Healthcare Worker Infection Table} of Hubei records the number of confirmed healthcare worker infections every day in each city of Hubei Province.

\begin{algorithm}[h]  
	\caption{Construction of Infection Table}  
	\begin{algorithmic}[1]
		\Require $Healthcare\ Infection\ Data =\{cms_i\}$
		\Ensure $Table^{Wuhan}(S,T)$
		\State $Table^{Wuhan}(S,T)=0$ 
		\For{each $cms \in \{cms_i\}$}  
		\State $Table^{Wuhan}(cms.County,cms.Date)+=1$ 
		\EndFor \\
		\Return $Table^{Wuhan}$
	\end{algorithmic}
\end{algorithm}

\subsection{Temporal and Spatial Characteristics Index}
The \emph{Healthcare Worker Infection Table} and the \emph{ Patient Infection Table} are essentially a collection of time series that contain spatial location information. Therefore, we further analyzed the temporal and spatial distribution characteristics of healthcare worker infections and patient infections.

In terms of temporal characteristics, we used eight indicators, namely the mean, standard variance, peak, rising period, rising rate, falling period, falling rate, and inflection point, to study the temporal characteristics of healthcare worker and patient infections in China (excluding Hubei), Hubei (excluding Wuhan), and Wuhan. In the Chinese scale, the main reasons for excluding Hubei Province are as follows: compared with other provinces, the number of patients and healthcare worker infections in Hubei Province is extremely high; if Hubei Province is not excluded, the temporal characteristics of China are only determined by the data of Hubei. Similarly, in the Chinese scale, Wuhan city is also excluded.

In terms of spatial characteristics, we divided the \emph{Healthcare Worker Infection Table } and the \emph{Patient Infection Table } based on time and used thematic maps to show the spatial distribution characteristics of healthcare worker and patient infections at each time node from China, Hubei, and Wuhan.

We determined the spatiotemporal differences between infections in healthcare workers and patients by assessing the differences in indices. A relatively small difference in indices is indicative of a relatively small spatiotemporal difference between infections in healthcare workers and patients, and vice versa.

\subsection{Pearson Correlation Coefficient}

The Pearson correlation coefficient \cite{b30} was calculated to quantitatively measure the spatiotemporal correlation between healthcare worker infections and patient infections, i.e., the difference in space and time between healthcare worker infections and patient infections. As shown in Figure 5, the spatial correlation was obtained by $t_3^h$ and $t_3^p$, and the temporal correlation was obtained by $s_3^h$ and $s_3^p$ \cite{b31}. The temporal and spatial correlation coefficients of the two series were calculated using Equations (1) and (2), respectively:

\begin{equation}
r_{temporal}  = \frac{{{\rm{Cov(}}s_i^p {\rm{,}}s_i^h {\rm{)}}}}{{\sqrt {\rm{D}(s_i^p) } \sqrt {\rm{D}(s_i^h)} }}
\end{equation} 

\begin{equation}
r_{spatial}  = \frac{{{\rm{Cov(}}t_i^p {\rm{,}}t_i^h {\rm{)}}}}{{\sqrt {\rm{D}(t_i^p )} \sqrt {\rm{D}(t_i^h)} }}
\end{equation} 
where $\rm{Cov}(X,Y)$ is the covariance of random variables $X$ and $Y$, $\rm{D}(X)$ is the variance of the random variable $X$, and $\sqrt{\rm{D}(X)}$ is the standard deviation of the random variable $X$. The value range of correlation coefficient $r$ is $[-1,1]$. When $r$>0, the random variables $X$ and $Y$ are positive correlations, if $r$=1, random variables $X$ and $Y$ are completely positively correlated; when $r$<0, the random variables $X$ and $Y$ are negatively correlated, if $r$=-1, then the random variables $X$ and $Y$ are completely negatively correlated. To measure the time correlation between healthcare worker infections and patient infections, $X$ and $Y$, respectively, represent the time series of healthcare worker infections and patient infections in the same study object. Similarly, to measure the spatial correlation between healthcare workers and patients, $X$ and $Y$, respectively, represent the number of healthcare worker infections and patient infections in multiple regions at a specific time. A strong positive correlation between two random variables is indicative of a relatively small spatiotemporal difference between infections in healthcare workers and patients, and vice versa.

\Figure[t!](topskip=0pt, botskip=0pt, midskip=0pt)[width=15.0cm]{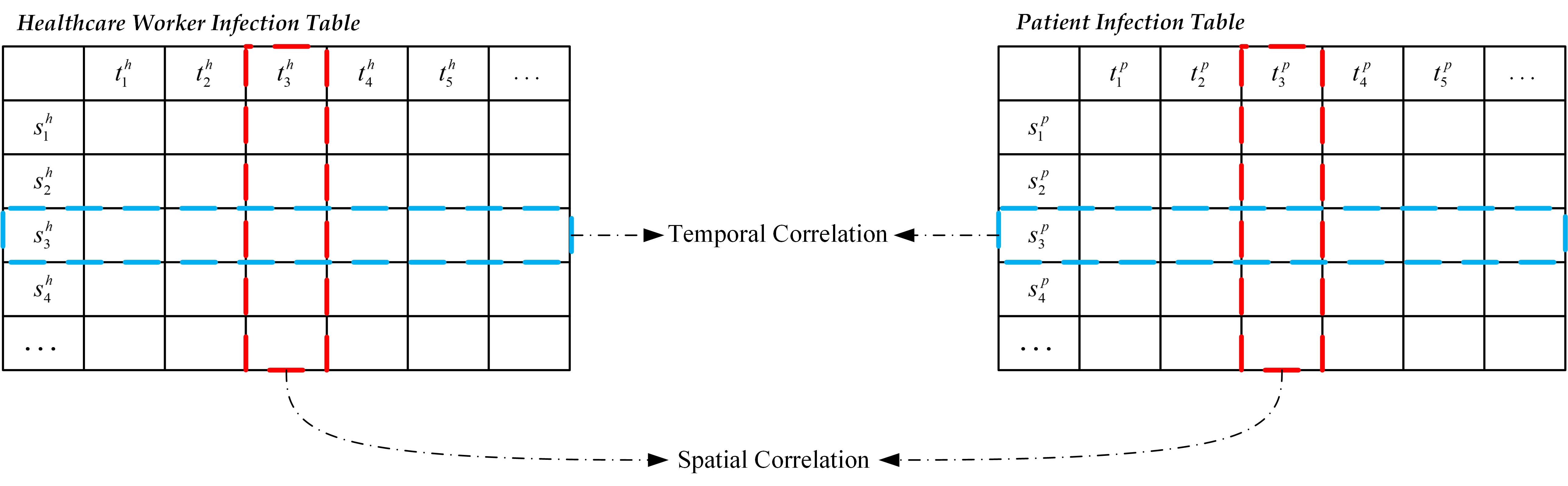}
{Temporal and spatial correlation coefficients.\label{fig5}}
\section{Experimental Results and Analysis}
\subsection{Temporal Differences between Healthcare Worker and Patient Infections}

\Figure[t!](topskip=0pt, botskip=0pt, midskip=0pt)[width=17.0cm]{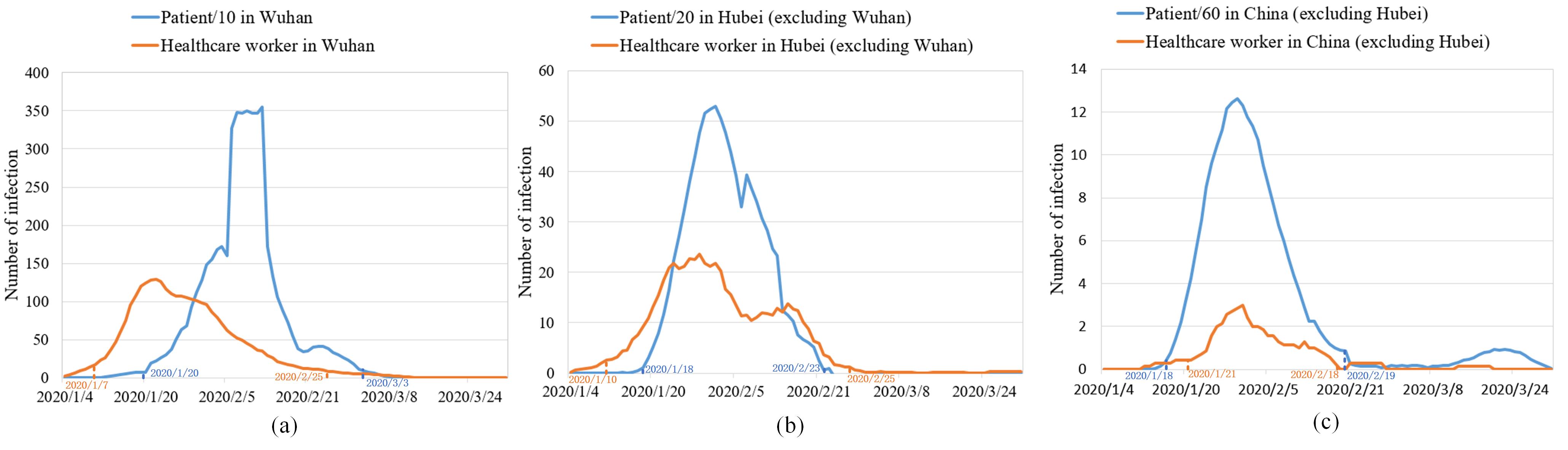}
{Comparison between the temporal characteristics of healthcare worker infection data and patient infection data: (a) Wuhan City, (b) Hubei Province excluding Wuhan, and (c) China excluding Hubei.\label{fig6}}

In this study, the statistical graph was first used to qualitatively describe the changes in the number of daily healthcare worker infections and patient infections over time. To make the curve smoother, we used the moving average method using a time window of seven days. The results are shown in Figure 6. Overall, the curves of daily healthcare workers infection and daily patients infection in China (excluding Hubei), Hubei Province (excluding Wuhan), and Wuhan City all initially showed a rising trend followed by a falling one. At the same time, the peaks of infections in China (excluding Hubei), Hubei (excluding Wuhan), and Wuhan have successively decreased. However, the curves of confirmed healthcare workers and confirmed patients were significantly different over time. For example, Figure 6(a) shows the onset of infection in healthcare workers, while it was absent in patients during the first week. There are two main reasons for this difference. First, the statistical approaches of the two data sources are different, which leads to the difference between the data. Second, and more importantly, data of healthcare worker infections were retrospective, while patient infections data were reported in real time. 

In addition, there were also significant differences in the statistical characteristics of the curves for healthcare worker and patient infections (Table 2). For example, The mean times of onset of infection for healthcare workers in Wuhan, Hubei (excluding Wuhan), and China (excluding Hubei) were January 27, February 1, and February 2, 2020, while those for patients were February 8, February 2, and January 31, 2020, the differences being 12, 1, and 2 days, respectively. The standard deviation values of infection onset time for healthcare workers in Wuhan, Hubei (excluding Wuhan), and China (excluding Hubei) were 9, 10, and 6 days, while those for patients were 7, 7, and 6 days, the differences being 2, 3, and 0 days, respectively. The results showed the mean time of onset for healthcare workers infection in Wuhan and Hubei (except Wuhan) was earlier than those for patients infection, and the mean times of onset for healthcare workers infection in China (except Hubei) was later than those for patients infection. There are two main reasons. First, the epidemics in Wuhan and Hubei (except Wuhan) are more serious. Compared with patient infections, healthcare worker infections are more likely to be detected. Second, healthcare workers have professional protective measures, so it is easier to control the spread of the epidemic among healthcare workers.

In addition, Pearson’s correlation coefficient was also used to measure the temporal correlation between healthcare worker and patient infections (Figure 7). The correlation coefficient r values for Wuhan, Hubei (excluding Wuhan), and China (excluding Hubei) were 0.26, 0.852, and 0.95, respectively. There are two main reasons for the lowest temporal correlation coefficient r in Wuhan. First, in the early stage of the epidemic, the protection of healthcare workers was insufficient, which made it easy to cause large-scale infections in hospitals. Second, compared with patient infections, infections in healthcare workers are more easily detected, leading to them being counted. The correlation coefficient also confirms a temporal difference between healthcare worker and patient infections.

In general, healthcare worker infection is contained earlier and faster. The temporal difference in Wuhan was greater than that in Hubei (excluding Wuhan), and the temporal difference in Hubei (excluding Wuhan) was greater than that in China (excluding Hubei). Combined with the time of epidemic occurrence in different regions, the temporal difference in the early phase was greater than that in the later phase, and the temporal difference in the early-onset region was greater than that in the later-onset region.

\begin{table}[]
	\caption{Statistical characteristics.}
	\begin{tabular}{cccc}
		\hline
		Index & Study area & Healthcare Worker & Patient \\ \hline
		\multirow{3}{*}{Mean}                                                         & Wuhan                & Jan 27            & Feb 8   \\ 
		& Hubei(outside Wuhan) & Feb 1             & Feb 2   \\  
		& China(outside Wuhan) & Feb 2             & Jan 31  \\ \hline
		\multirow{3}{*}{\begin{tabular}[c]{@{}c@{}}Standard\\ deviation\end{tabular}} & Wuhan                & 9 days            & 7 days  \\ 
		& Hubei(outside Wuhan) & 10 days           & 7 days  \\ 
		& China(outside Wuhan) & 6 days            & 6 days  \\ \hline
		\multirow{3}{*}{\begin{tabular}[c]{@{}c@{}}Inflection\\ point\end{tabular}}   & Wuhan                & Jan 22            & Feb 12  \\ 
		& Hubei(outside Wuhan) & Jan 29            & Feb 1   \\ 
		& China(outside Wuhan) & Jan 30            & Jan 31  \\ \hline
		\multirow{3}{*}{Peak}                                                         & Wuhan                & 128               & 3550    \\  
		& Hubei(outside Wuhan) & 23                & 1058    \\ 
		& China(outside Wuhan) & 3                 & 756     \\ \hline
		\multirow{3}{*}{\begin{tabular}[c]{@{}c@{}}Rising\\ period\end{tabular}}      & Wuhan                & 15 days           & 23 days \\ 
		& Hubei(outside Wuhan) & 19 days           & 14 days \\ 
		& China(outside Wuhan) & 12 days           & 10 days \\ \hline
		\multirow{3}{*}{\begin{tabular}[c]{@{}c@{}}Rising\\ rate\end{tabular}}        & Wuhan                & 64                & 1617    \\  
		& Hubei(outside Wuhan) & 12                & 550     \\ 
		& China(outside Wuhan) & 1                 & 464     \\ \hline
		\multirow{3}{*}{\begin{tabular}[c]{@{}c@{}}Falling\\ rate\end{tabular}}       & Wuhan                & 34 days           & 20 days \\  
		& Hubei(outside Wuhan) & 27 days           & 22 days \\ 
		& China(outside Wuhan) & 19 days           & 19 days \\ \hline
		\multirow{3}{*}{\begin{tabular}[c]{@{}c@{}}Falling\\ rate\end{tabular}}       & Wuhan                & 55                & 525     \\ 
		& Hubei(outside Wuhan) & 11                & 450     \\ 
		& China(outside Wuhan) & 1                 & 334     \\ \hline
	\end{tabular}
\end{table}

\Figure[h!](topskip=0pt, botskip=0pt, midskip=0pt)[width=14.0cm]{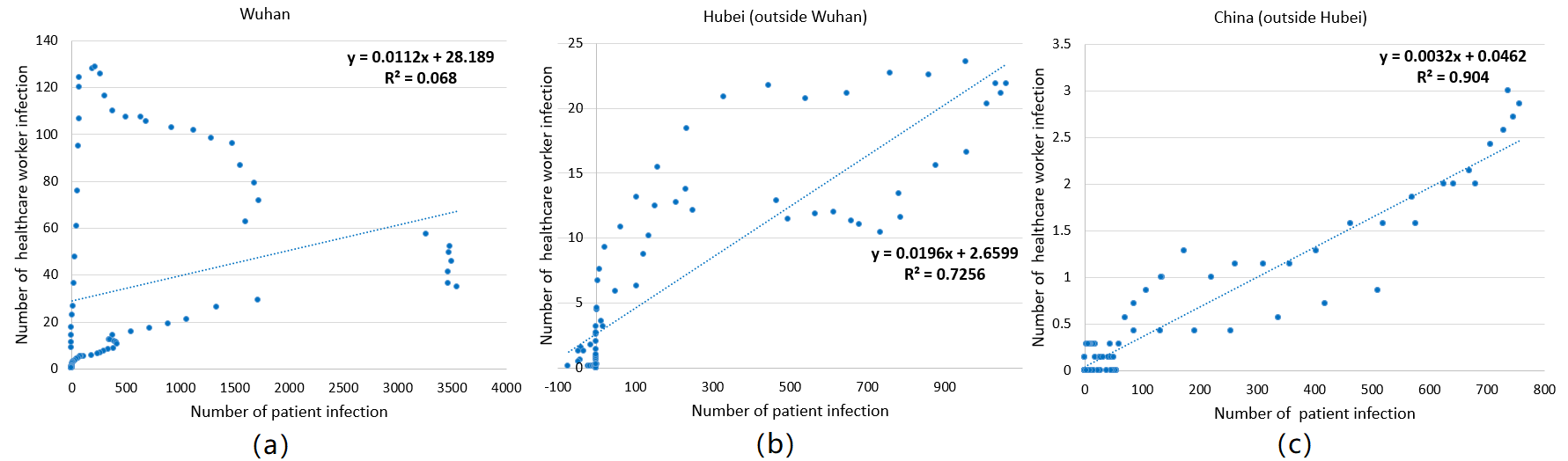}
{Temporal correlations of healthcare worker infection data and patient infection data: (a) Wuhan City, (b) Hubei Province excluding Wuhan, and (c) China excluding Hubei.\label{fig7}}
\subsection{Spatial Differences between Healthcare Worker Infection and Patient Infection}
The spread of COVID-19 in space is a dynamic process. To further compare the spatial distribution characteristics and differences between healthcare worker and patient infections, we first analyzed the spatial distribution characteristics of healthcare worker and patient infections at each time node with a time interval of 10 days and then quantitatively determined the spatial correlation between healthcare worker and patient infections at each time node. In this section, the period from January 7, 2020, to February 16, 2020 was selected as the research interval. 

\subsubsection{Spatial Characteristics of Healthcare Worker Infection}

Figure 8 shows the spatial distribution of healthcare worker infections in China, Hubei, and Wuhan. In China, healthcare worker infections first occurred in Hubei, and then the phenomenon of healthcare worker infection gradually spread outwards through Hubei. As of February 6, 2020, the number of healthcare worker infections in different provinces of China significantly differed. Generally, the provinces with many healthcare worker infections in China were mainly located in the southeast coastal areas of China; these provinces are close to Hubei Province and have better economic development. The provinces with relatively low healthcare worker infections were mainly located in northwestern China; these provinces are far from Hubei and economically underdeveloped.

In Hubei, healthcare worker infections first occurred in Wuhan and Huanggang and then spread across, with Wuhan and Huanggang essentially being the center. As of January 27, 2020, most of the cities in Hubei Province had different degrees of healthcare worker infections, among which Wuhan and Huanggang had more more severe infections. Generally, cities with more severe healthcare worker infections are mainly concentrated in the eastern part of Hubei province, while the cities with fewer healthcare worker infections are mainly concentrated in the western part of Hubei Province.

In Wuhan, healthcare worker infections first occurred in Wuhan Jianghan District and Jiangan District and then spread to the two adjacent districts. As of February 06, 2020, all districts in Wuhan had different degrees of healthcare worker infection, among which the districts with more severe healthcare worker infections mainly occurred in the central district of Wuhan, such as Jianghan and Wuchang, where the number of healthcare worker infections exceeded 500. The districts with mild healthcare worker infections were mainly located in Jiangxia and Huangpo outside Wuhan, but there were still more than 10 cases of healthcare worker infections. In general, healthcare worker infections in Wuhan were severe.

\Figure[t!](topskip=0pt, botskip=0pt, midskip=0pt)[width=14.5cm]{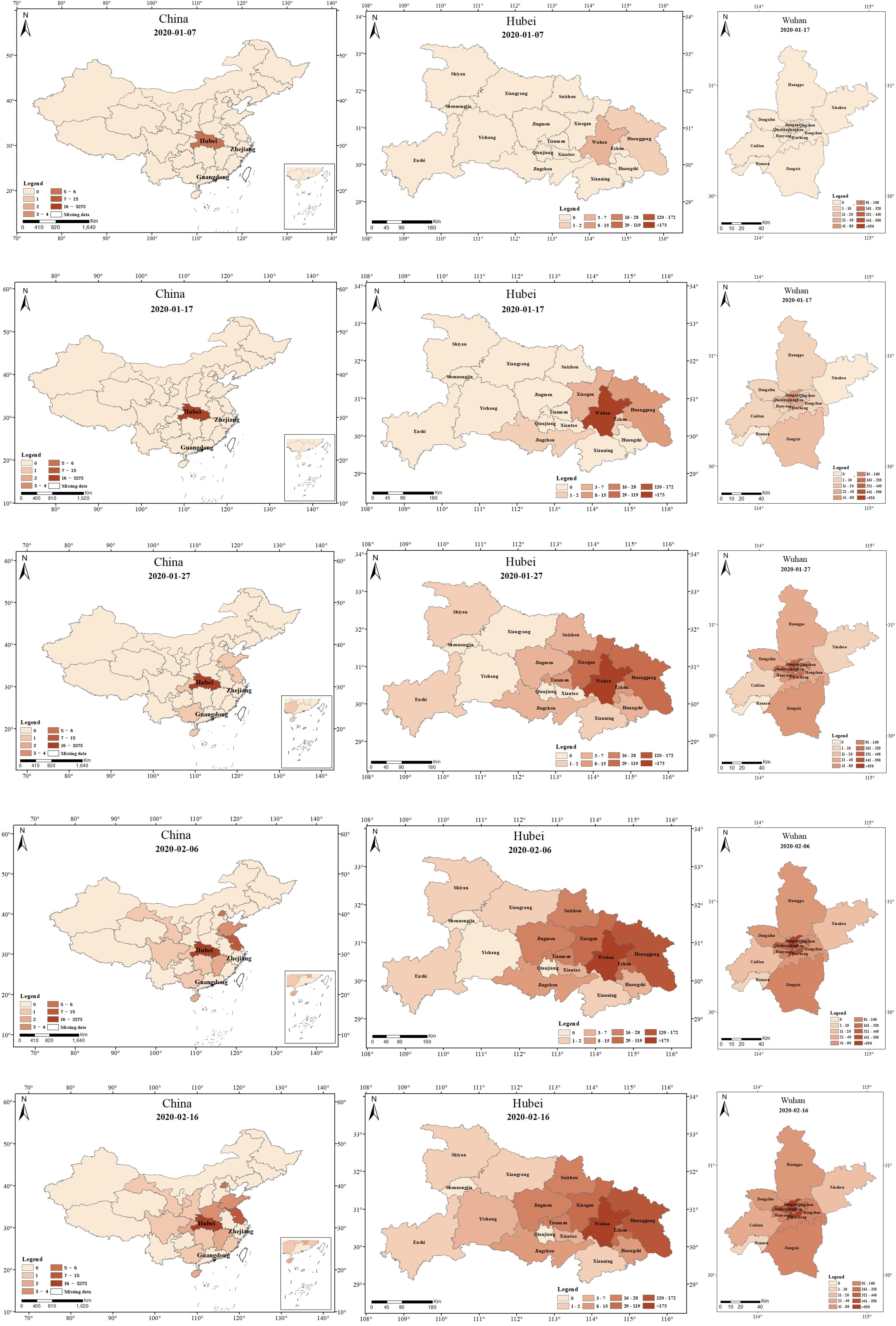}
{Spatial distribution characteristics of healthcare worker infections at different time nodes in China, Hubei, and Wuhan in 2020: (a) on January 7, (b) on January 17, (c) on January 27, (d) on February 6, (e) on February 16. \label{fig8}}

\subsubsection{Spatial Characteristics of Patient Infection}
As a comparison with the spatial distribution characteristics of healthcare infection, we showed the spatial distribution characteristics of patient infection during the same period. As it is difficult to obtain data on patient infections at the county level in Wuhan from mid-January to mid-February, we showed the spatial distribution characteristics of patient infection only at the scale of China and Hubei. The results are shown in Figure 9.

\Figure[t!](topskip=0pt, botskip=0pt, midskip=0pt)[width=15.cm]{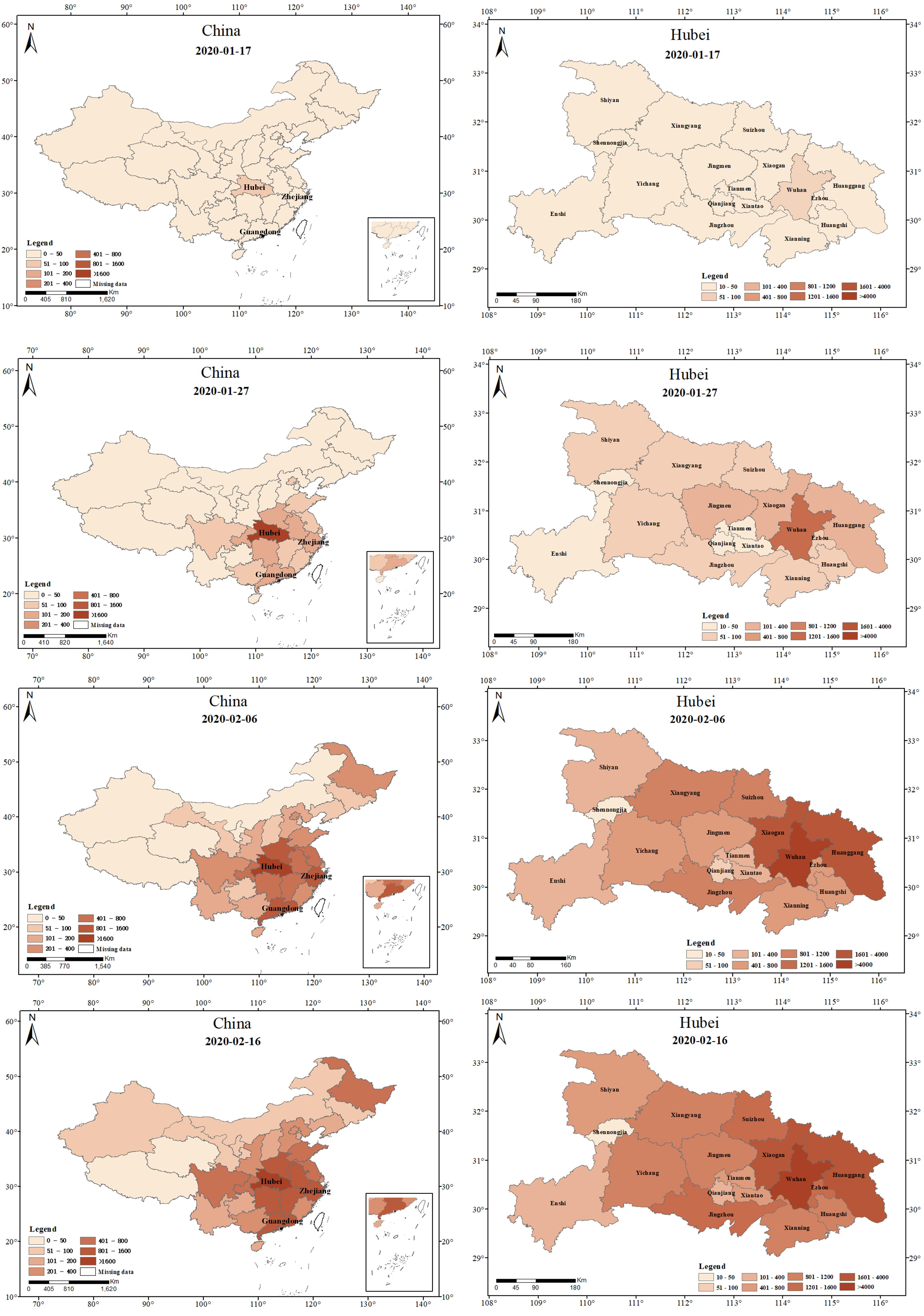}
{Spatial distribution characteristics of patient infections at different time nodes in China and Hubei in 2020: (a) on January 17, (b) on January 27, (c) on February 6, (d) on February 16.\label{fig9}}

In China, the first case of infection in a patient was reported in Hubei and then spread outward through Hubei. As of February 06, 2020, the numbers of patient infections in different provinces of China were significantly different. Compared with healthcare worker infections, in addition to the difference in the number of infections, there were also certain differences in the spatial distribution of infection. For example, the patient infections in Guangdong and Zhejiang were relatively serious, but the healthcare infection in the two provinces were relatively mild. This may be attributable to the differences in the epidemic prevention strategies of different provinces. For example, Guangdong Province acquired more valuable experience during the SARS outbreak and thus managed to swiftly protect healthcare workers.

In Hubei, the first case of infection in a patient was reported in Wuhan City and then spread outward throughout Wuhan City. As of February 06, 2020, the spatial difference was relatively small, except for the large difference in the number of infections. In other words, cities with relatively more severe patient infections have relatively more severe healthcare infections, and cities with less patient infections have less healthcare infections. 

Further, to measure the degree of correlation quantitatively between healthcare worker infection and patient infection, we analyzed the spatial correlation between healthcare worker and patient infections (Table 3). The correlation coefficients, r, for China (except Hubei) on January 27, February 06, and February 16 were 0.132, 0.135, and 0.252 respectively, indicating that the spatial distributions of healthcare worker and patient infections in China (except Hubei) are quite different. The correlation coefficients, r, for Hubei (except Wuhan) on January 27, February 06, and February 16 were 0.845, 0.816, and 0.859, respectively, which indicated that the spatial distribution difference between healthcare worker and patient infections in Hubei (except Wuhan) was small.

In general, in addition to the differences in the number of healthcare worker and patient infections, the spatial distribution was different to some extent. The spatial distribution difference in Hubei (except Wuhan) was greater than that in China (except Hubei), that is, the difference in the late stage was greater than that in the early stage; the difference in late-onset regions was greater than that in the early-onset regions.

\begin{table}[]
\caption{Spatial correlations of healthcare worker and patient infections in China (excluding Hubei) and Hubei (excluding Wuhan).}
\begin{tabular}{lcc}
\hline
\multicolumn{1}{c}{Date}      & China (outside Hubei) & Hubei (outside Wuhan) \\ \hline
\multicolumn{1}{c}{2020-01-27} & $r$= 0.132              & $r$= 0.845              \\ 
2020-02-06                       & $r$= 0.159              & $r$= 0.816              \\ 
2020-02-16                       & $r$= 0.252              & $r$= 0.859              \\ \hline
\end{tabular}
\end{table}

\subsection{Limitations of the Study}

The limitations of this study are as follows: (1) the coverage of healthcare worker infection data was narrow, and this study only examined the temporal and spatial distributions and differences between healthcare worker and patient infections in China. Owing to the difficulty in obtaining data on healthcare worker infections in other countries, the temporal and spatial distributions and differences in other countries were not compared and analyzed. (2) The infection data of healthcare workers were not comprehensive. As the information on healthcare worker infections is published on the official website of the Red Cross Foundation in batches, information may continue to be published in the future; thus, this study may be missing information. (3) In terms of the spatial distribution, it is difficult to obtain patient infection data at the county level in Wuhan for the same period. This study only examined the spatial distributions and differences in China (excluding Hubei) and Hubei (excluding Wuhan). In response to the above problems, future research direction will focus on further collecting domestic and foreign healthcare workers and patient infection data to analyze the temporal and spatial distributions and differences more accurately and comprehensively between healthcare worker and patient infections.

\section{Discussions and Conclusion}
At present, COVID-19 is still a pandemic. As specific vaccines and fundamental treatments take a long time, relevant measures such as self-isolation, restriction of crowd activities and gathering, and wearing of masks have become important emergency containment measures. To accurately formulate epidemic containment policies, it is important to study the temporal and spatial spread of COVID-19 in social groups.

This study analyzed the temporal and spatial spread characteristics of and differences in COVID-19 among patients and healthcare workers from three perspectives: China (excluding Hubei), Hubei (excluding Wuhan), and Wuhan. 

In terms of temporal characteristics, healthcare worker and patient infections in Wuhan, Hubei (excluding Wuhan), and China (excluding Hubei) initially tended to increase and then decrease; the peak value of infection gradually decreased. However, there are great differences between the two curves. For example, the correlation coefficients, r, between healthcare worker and patient infections for Wuhan, Hubei (excluding Wuhan), and China (excluding Hubei) were 0.26, 0.852, and 0.95, respectively. In other words, the temporal differences between healthcare worker and patient infections were relatively large in Wuhan, whereas they decreased successively in Hubei (excluding Wuhan) and China (excluding Hubei). Combined with the time of the occurrence of the epidemic in different regions, the temporal difference in the early stage is greater than that in the later stage between healthcare worker and patient infections, and the temporal difference in the early-onset region is greater than that in the late-onset region. Through temporal difference, different policies can be formulated for different social groups through the following aspects. For example, it is very important to strengthen the protective measures for healthcare workers in the early stage of the epidemic. Because healthcare workers work in a relatively closed space, the battle is on the frontline of the epidemic. Even if the number of patients infected is small, it can still result in many infections among healthcare workers. In addition, we also can use the temporal difference to predict the peak time of infection in patients \cite{b32}.

In terms of spatial characteristics, healthcare worker and patient infections in China spread mainly through Hubei as the epicenter. Healthcare worker and patient infections in Hubei spread mainly through Wuhan and Huanggang as the epicenter. Healthcare worker infection and patient infections in Wuhan spread mainly through Jianghan and Jiangan as the epicenter. In addition, the spatial distributions of healthcare worker and patient infections also differed from different scales. For example, on January 27, February 06, and February 16, 2020, the correlation coefficients, r, between healthcare worker infection and patient infection for China (excluding Hubei) were 0.132, 0.135, and 0.252, respectively. Further, the correlation coefficients, r, between healthcare worker infection and patient infection in Hubei (excluding Wuhan) were 0.845, 0.816, and 0.859, respectively. Combined with the time of epidemic occurrence in different regions, the spatial difference between healthcare worker infection and patient infections in the early stage was less than that in the later stage, and the spatial difference in the early-onset region was less than that in the late-onset region between healthcare worker and patient infections. The results of this study offer significant guidance for epidemic containment departments to accurately formulate different epidemic containment policies for different social groups. Through spatial correlation, we can use the spatial distribution of healthcare worker infection to infer the spatial distribution of patient infection in the early stage of the epidemic.

\bibliographystyle{IEEEtran}
\bibliography{access}

\begin{IEEEbiography}[{\includegraphics[width=1in,height=1.25in,clip,keepaspectratio]{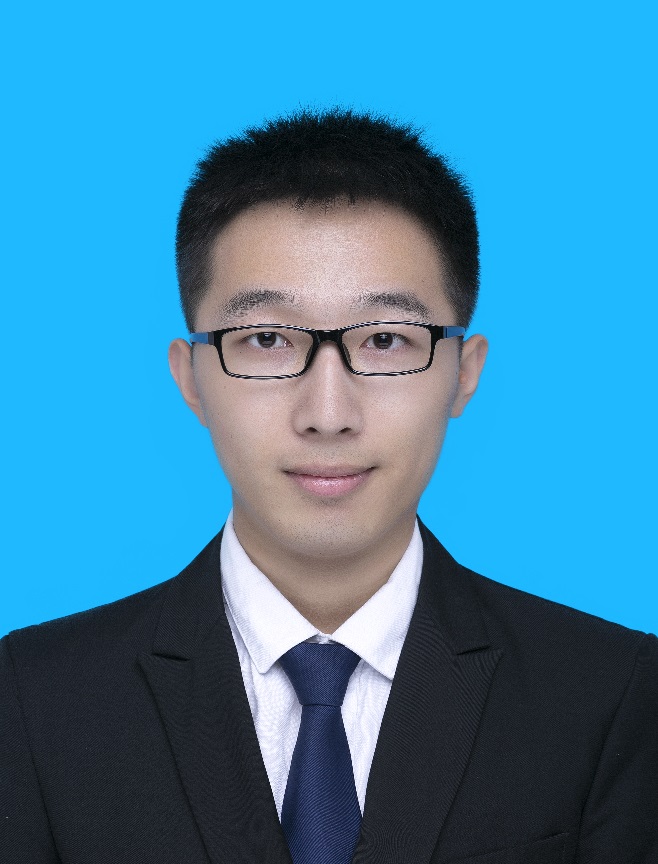}}]{PEIXIAO WANG} is a PhD candidate of State Key Laboratory of Information Engineering in Surveying, Mapping and Remote Sensing, Wuhan University. His research focus on geographic information service and spatio-temporal data mining. Email: peixiaowang@whu.edu.cn.
\end{IEEEbiography}
\begin{IEEEbiography}[{\includegraphics[width=1in,height=1.25in,clip,keepaspectratio]{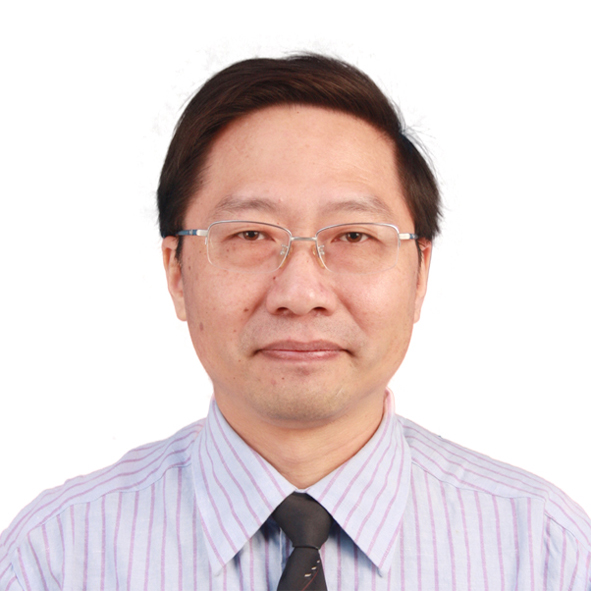}}]{XINYAN ZHU} received the Ph.D. degree from the State Key Laboratory of Information Engineering in Surveying, Mapping and Remote Sensing, Wuhan University.He is currently a Professor with Wuhan University. During the past years, he has published over 100 refereed journal articles and conference papers. His research interests cover spatio-temporal ata modeling and analysis,distributed spatial database, holographic position map and its application, indoor location and navigation, social geographic computing, and public health.Email:xinyanzhu@whu.edu.cn
\end{IEEEbiography}

\begin{IEEEbiography}[{\includegraphics[width=1in,height=1.25in,clip,keepaspectratio]{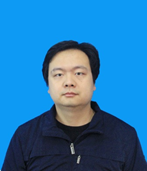}}]{WEI GUO} received the Ph.D. degree from the State Key Laboratory of Information Engineering in Surveying, Mapping and Remote Sensing, Wuhan University.He is currently a Associate Professor with Wuhan University. During the past years, he has published over 40 refereed journal articles and conference papers. His research interests cover spatio-temporal ata modeling and analysis,distributed spatial database,and public health.Email:guowei-lmars@whu.edu.cn
\end{IEEEbiography}

\begin{IEEEbiography}[{\includegraphics[width=1in,height=1.25in,clip,keepaspectratio]{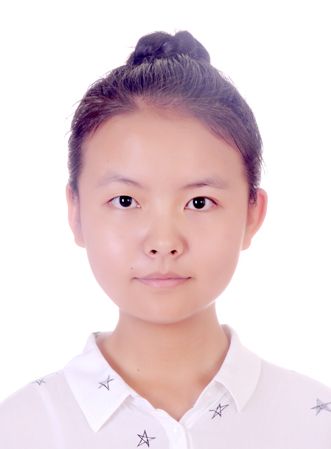}}]{HUI REN} is currently pursuing the M.S. degree with State Key Laboratory of Information Engineering in Surverying,Mapping and Remote Sensing,Wuhan University. Her research focus on public health and spatiotemporal analysis. Email:renhui@whu.edu.cn
\end{IEEEbiography}

\begin{IEEEbiography}[{\includegraphics[width=1in,height=1.25in,clip,keepaspectratio]{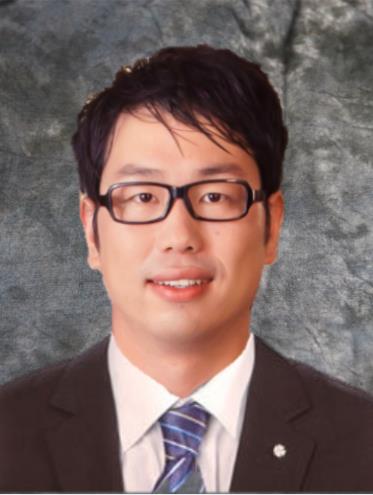}}]{TAO HU} is a post-doctoral fellow at Harvard Center for Geographic since July 2019. He is working on the Spatial Data Lab project. Before that, he was a post-doctor researcher at Kent State University since 2016. He received Ph.D. and B.S in GIS from Wuhan University. His primary research field is GIScience, with a focus on geospatial data analytics, spatiotemporal analysis with applications to urban crimes, social sensing, and public health. He has more than 20 peer-reviewed publications and serves as a peer reviewer for more than 10 international journals. Email:xinyanzhu@whu.edu.cn
\end{IEEEbiography}

\EOD

\end{document}